# The maximal number of degenerate directions for non-piezoelectric media of trigonal symmetry

Artur Duda, Institute of Theoretical Physics,
University of Wroclaw, pl. Maxa Borna 9,
52-204,Wroclaw, Poland.

## Abstract
We show that one can construct positively defined matrix of elastic constants representing medium of trigonal symmetry for which exactly 16 distinct degenerate directions exist.

Degenerate directions are normals along which sound waves with at least two modes possessing equal phase velocity could propagate. These normals are also sometimes called acoustic axes or singularities. Conditions for effective phonon interaction are fulfilled, as a rule, in the vicinity of degenerate direction. Their number and orientation have been the subject of research for many years.

The number and orientation of degenerate directions for media of trigonal symmetry were firstly investigated by Khatkevich [1]. He found that crystals of trigonal symmetry could have 4 or 10 degenerate directions and that all these directions are situated inside symmetry planes of the crystal.

Trigonal crystals are main substrate materials for surface acoustic wave devices which find widespread application in modern signal processing equipment. This motivated recent thorough study of degenerate directions for crystals of trigonal symmetry conducted by Mozhaev et al [2-3]. They found that Khatkevich's statement about the absence of degenerate directions situated outside symmetry planes isn't correct and derived expressions for spherical coordinates of these directions. The existence of oblique degenerate directions in principle could increase their total maximal number to 16. Nevertheless computational experiments performed by Mozhaev et al haven't shown that the total number of degenerate directions for nonpiezoelectric media of trigonal symmetry could increase due to the existence of degenerate directions situated outside symmetry planes. They found that for sapphire the artificial increase of the value of the elastic constant $C_{14}$ caused the replacement of degenerate directions in the symmetry planes by degenerate directions out of symmetry planes.

We will give below an example of the matrix of elastic constants for which 16 distinct degenerate directions exist. Let us consider the medium described by the following matrix of elastic constants

$$C = \begin{pmatrix} 157.4 & 10.43 & 11.91 & -27.91 & 0 & 0 \\ 10.43 & 157.4 & 11.91 & 27.91 & 0 & 0 \\ 11.91 & 11.91 & 130.5 & 0 & 0 & 0 \\ -27.91 & 27.91 & 0 & 157.3 & 0 & 0 \\ 0 & 0 & 0 & 0 & 157.3 & -27.91 \\ 0 & 0 & 0 & 0 & -27.91 & 73.485 \end{pmatrix}$$

Degenerate directions characterizing above medium are collected in the below given table.

| Type of degenerate directions | Components of degenerate directions |
|---|---|
| Degenerate directions situated inside symmetry planes | [0, 0.864, 0.503]; [0, -0.118, 0.993]; [0, -0.919, 0.394] [-0.744, -0.439, 0.503]; [0.102; 0.059; 0.993]; [0.795; 0.461; 0.394]; [0.744; -0.439; 0.503]; [-0.102; 0.059; 0.993]; [-0.795; 0.461; 0.394] |
| Oblique degenerate direction | [-0.862; -0.143; 0.486]; [0.862; -0.143; 0.486]; [-0.312; 0.817; 0.486]; [0.559; -0.671; 0.486]; [0.312; 0.817; 0.486]; [-0.559; -0.671; 0.486] |
| Threefold symmetry axis | [0, 0, 1] |

The total number of degenerate directions is 16.